# Magnetic field dependent direct rf power absorption studies in $Bi_2Sr_2CaCu_2O_8$ single crystals

S. P. Chockalingam, S. Sarangi and S. V. Bhat[1]

Department of Physics, Indian Institute of Science, Bangalore-560012, India

K. Oka* and Y. Nishihara**

*National Institute of Advanced Industrial Science and Technology, Tsukuba, 310-8512, Japan;   **Faculty of Science, Ibaraki University, Mito 310-8512, Japan

[1] Corresponding author:

Prof. S.V.Bhat,
Department of Physics
Indian Institute of Science
Bangalore –560012, India
e-mail: svbhat@physics.iisc.ernet.in
Tel.: +91-80-22932727. Fax: +91-80-23602602

# Abstract:

Using a novel technique of measuring direct rf power dissipation, the temperature and magnetic field dependence of rf absorption in superconducting single crystals of $Bi_2Sr_2CaCu_2O_8$ is studied. An unexpected, larger than the normal state, dissipation is observed below $T_c$, which exhibits peaks at magnetic field dependent temperatures. An explanation is provided for this observation in terms of the energy required to decouple the intrinsic Josephson junctions in the crystals.

**Introduction:**

Magnetic field dependent ac (rf and microwave) losses in high $T_c$ superconductors have been studied extensively in the last few years [1]. Such studies not only yield information on the material parameters crucial for applications but can also provide a test for any possible microscopic theory of superconductivity. By now it is established that ac power dissipation in high $T_c$ superconductors can arise mainly due to two reasons; the decoupling of Josephson junctions (JJ) (which are invariably present in the high $T_c$ materials, either as extrinsic junctions due to intergranular contacts or as intrinsic junctions in strongly layered superconductors like $Bi_2Sr_2CaCu_2O_8$) and due to Lorentz force driven motion of the vortices (Josephson and pancake) in these strongly type II superconductors [2]. In both these cases, it is expected and mostly observed that the power absorption decreases with decreasing temperature and increases with increasing magnetic field. However, over the years a large number of examples has also been found where the dependence of the power absorption on the field is anomalous i.e. the power absorbed decreases with increasing field [3]. A number of possible explanations such as the Paramagnetic Meissner Effect (PME) or Wohllben effect [4], the d-wave nature of the super conducting wave function [5], a percolation mechanism in degraded samples [6] and the distribution of junction strengths in superconducting loops [7] have been proposed to understand such anomalous field dependence of absorption.

Most of the earlier studies of ac dissipation in high $T_c$ materials have used the technique of non-resonant microwave absorption [NRMA] or non-resonant rf absorption [NRRA], where the field derivative of the power absorbed, dP/dH, is measured. It is desirable to use a technique where one can measure the power absorbed, P(H,T), directly as a function of magnetic field and temperature in the superconducting state. Recently we have designed and fabricated [8] such an apparatus using which we have studied a number of superconducting samples. In some samples we have observed strongly anomalous field and temperature dependence of the rf dissipation [9]. In the present work we report results on the direct rf power absorption from well-characterized single crystals of $Bi_2Sr_2CaCu_2O_8$, Bi-2212, as a function of field and temperature and understand it in terms of a new model proposed by us [9], which explains the power absorption as a consequence of energy required to decouple Josephson junctions.

**Experimental:**

The absolute power absorption studies are carried out using an rf oscillator (typical operating frequency: 8 MHz, rf amplitude 0.7 V peak to peak) designed and fabricated in our laboratory [8]. The system consists of a self-resonant LC tank circuit driven by a NOT gate. The samples under investigation are placed in the core of an inductive coil forming the LC circuit and the power absorption is determined from the measured change in the total current supplied to the oscillator. A customized low temperature insert is used to integrate the experiment with a commercial Oxford Instruments cryostat and temperature controller (4.2 K < T < 300 K) along with a Bruker electromagnet (0 < H < 1.4 T).

The power absorption studies described here have been made on air annealed single crystals with the composition $Bi_2Sr_2CaCu_2O_8$. The crystals were grown by traveling zone flux method. The crystals had a nominal $T_c$ of ~85 K with the transition width $\Delta T \sim 1K$ as determined by ac susceptibility measurements. For the power dissipation studies the crystal was oriented such that the c-axis made an angle of 45° with the direction of the applied field. This orientation results in both the pancake and Josephson vortices getting excited in the crystal.

## Results:

Fig.1 shows the absolute rf power absorbed by the sample at different fields as a function of temperature. In the normal state the power absorption decreases linearly with temperature as expected from the resistivity behavior of metals. Below $T_c \sim 85$ K, in the superconducting state, the power absorption increases with decreasing temperature going through a peak. In the absence of the field we observe an absorption peak around 75 K. With increasing field the amplitude and the width of this peak (peak A) are observed to decrease and the peak position is seen to shift towards lower temperatures. For $H \geq 50$ mT, a second peak (peak B) appears to emerge around 75 K. With increasing field, the intensity of this peak increases though the position in temperature shows very little dependence on magnetic field. In figure 2, we present the field dependence of the positions of the two peaks, where the different behaviors of the two peaks are clearly seen.

In figures 3 and 4 we present the field dependence of the absorbed power at different temperatures. Fig.3 covers the field range of 0 to 1 T whereas fig. 4 represents the detailed behavior at low field region (0 to 300 mT). We would like to point out three interesting features of the field dependence:

(1) At all temperatures $T < T_c$, for moderately high fields ($H \geq 400$ mT), the dissipation decreases with increasing field, in contrast with what is generally expected and reported.

(2) For low fields, ($H \leq 400$ mT) the absorption goes through a peak, whose position, intensity and width are very sensitively temperature dependent. Such behavior has earlier been observed in microwave range of frequencies [10].

(3) Centered at zero field (fig.4) there is another absorption peak. With increasing low fields this dissipation decreases as well.

**Discussion:**

We have earlier studied the dissipation behavior in Bi-2212 single crystals using the technique of NRMA [11] and NRRA [12]. The results were interpreted in terms of the response of the intrinsic Josephson junctions and the vortex lattice to the applied field and temperature. Since in both of these experiments field modulation and lock-in detection were used, only the first derivative of the absorbed power could be measured. Now with the ability to study the direct power absorption, we notice certain additional features of the phenomenon. Most interestingly the dissipation in the superconducting state is larger than that in the normal state immediately above $T_c$. We have observed this phenomenon in some other superconducting samples as well earlier [9] and have proposed the following model to explain the same; the Josephson junctions (whether intrinsic or extrinsic) have a characteristic coupling energy $E_J$ given by $E_J = \Phi_o I_c / 2\pi$, where $\Phi_o$ is the flux quantum and $I_c$ is the critical current of the junction. With the sample getting exposed to the rf field, an rf current is induced in it. When the induced rf current exceeds $I_c$ for a given T and H, the junction goes normal. This leads to two contributions to the net dissipation: (1) energy required to decouple the junction which is equal to $E_J$ and (2) the normal state loss of the junction once the junction is decoupled. The total power absorbed is the sum of these two energies multiplied by the frequency.

With this model, our results on Bi-2212 can be satisfactorily explained as follows: as is well known [13], in the highly anisotropic crystals of Bi-2212, the insulating layers sandwiched between the superconducting Cu-O planes act as intrinsic

Josephson junctions. Peak A in figs. 1 and 2 arises because of the decoupling of these intrinsic junctions. For zero applied magnetic field, these junctions are formed close to $T_c$. Initially, when the sample is cooled below $T_c$, the number of JJ's goes on increasing, which on being subject to $I_{rf} > I_c$ decouple and contribute to an increase in the dissipation. As the temperature is decreased further, $E_J$ of the junctions increases such that $I_{rf} < I_c$ and $I_{rf}$ is not able to decouple the junctions and therefore a decrease in the absorbed power is observed. For increasing values of the magnetic field, $I_c = I_{rf}$ occurs at lower temperatures making the peak A to occur at lower and lower temperatures. Similar reasoning could explain the field dependence of the absorption as well.

The behavior of peak B is quite different from that of peak A. Not only its position in temperature is almost independent of magnetic field, but also its magnitude increases with increasing magnetic field. Detailed analysis of this absorption peak in terms of Josephson and pancake vortices will be discussed in a future publication.

In summary, using a newly developed technique of direct rf power absorption we have observed an apparently unusual larger than the normal state dissipation in superconducting single crystals of Bi-2212. We provide a qualitatively satisfactory explanation of this phenomenon in terms of the decoupling of intrinsic Josephson junctions in these strongly layered material.

**Acknowledgements,** SVB thanks the University Grants Commission, India for funding this work.

# References:

[1] For example, see the various review articles in *"Studies of High Temperature superconductors"*, ed. A.V.Narlikar, Vol. **17,18** (1996), Nova Science publishers, New York.

[2] G.Blatter, V.B.Geshkenbein, A.I. Larkin and V.M.Vinokur.
    Rev. Mod. Phys. **66,** 1125 (1994)

[3] S.V.Bhat, *Studies of High Temperature superconductors,* ed. by A.V.Narlikar,
    **Vol.18**, pp. 241-259, (1996).

[4] W.Braunisch, N.Knauf, D.Wohlleben et.al. Phys. Rev. Lett., **68**,1908 (1992)

[5] M.Sigrist and T.M.Rice. Rev.Mod.Phys. **67**,503 (1995)

[6] S.V.Bhat, V.V. Srinivasu, N.kumar. Phys. Rev. B, **44**, 10121 (1991)

[7] A.Anand and S.V.Bhat Physica C **341-348,** 1671-1672, (2000)

[8] S.Sarangi and S.V.Bhat, cond-mat 0503542; Rev. Sci. Instrum., **76,** 023905 (2005)

[9] S.Sarangi, S.P.Chockalingam and S.V.Bhat. cond-mat (0503401)

[10] E.J.Pakulis and G.V.Chandrasekhar. PRB 39,808(1989)

[11] A.Rastogi, Y.S.Sudershan, S.V.Bhat, A.K.Grover, Y.Yamaguchi, K.Oka, Y.Nishihara, Phys.Rev. **B. 53,** 9366, (1996)

[12] Y.S.Sudershan, Amit Rastogi, S.V.Bhat, A.K.Grover, Y.Yamaguchi, K.Oka,
       Y.Nishihara,
       Physica **C, 297,** 253-261, (1998)

[13] R. Kleiner and P. Muller, Phys. Rev. B **49,** 1327 (1994)

**Figure Captions:**

Figure 1: Absorbed rf power as a function of temperature at various applied magnetic fields. To be noted are the decrease in the width of the peaks as well as their position in temperature with increasing magnetic field.

Figure 2: Peak position (in temperature) as a function of the magnetic field for the two peaks. The sensitive dependence of the position of peak A (specially for low fields) and near temperature independence of the position of peak B are being seen.

Figure 3: Field dependence of the absorbed power at different temperatures. For high fields the dissipation is a decreasing function of the field at all temperatures, where as for low fields, a temperature dependent peak is observed.

Figure 4: Same as for fig.3 but for a smaller field range (0 to 300 mT) to show the peak behavior more clearly.

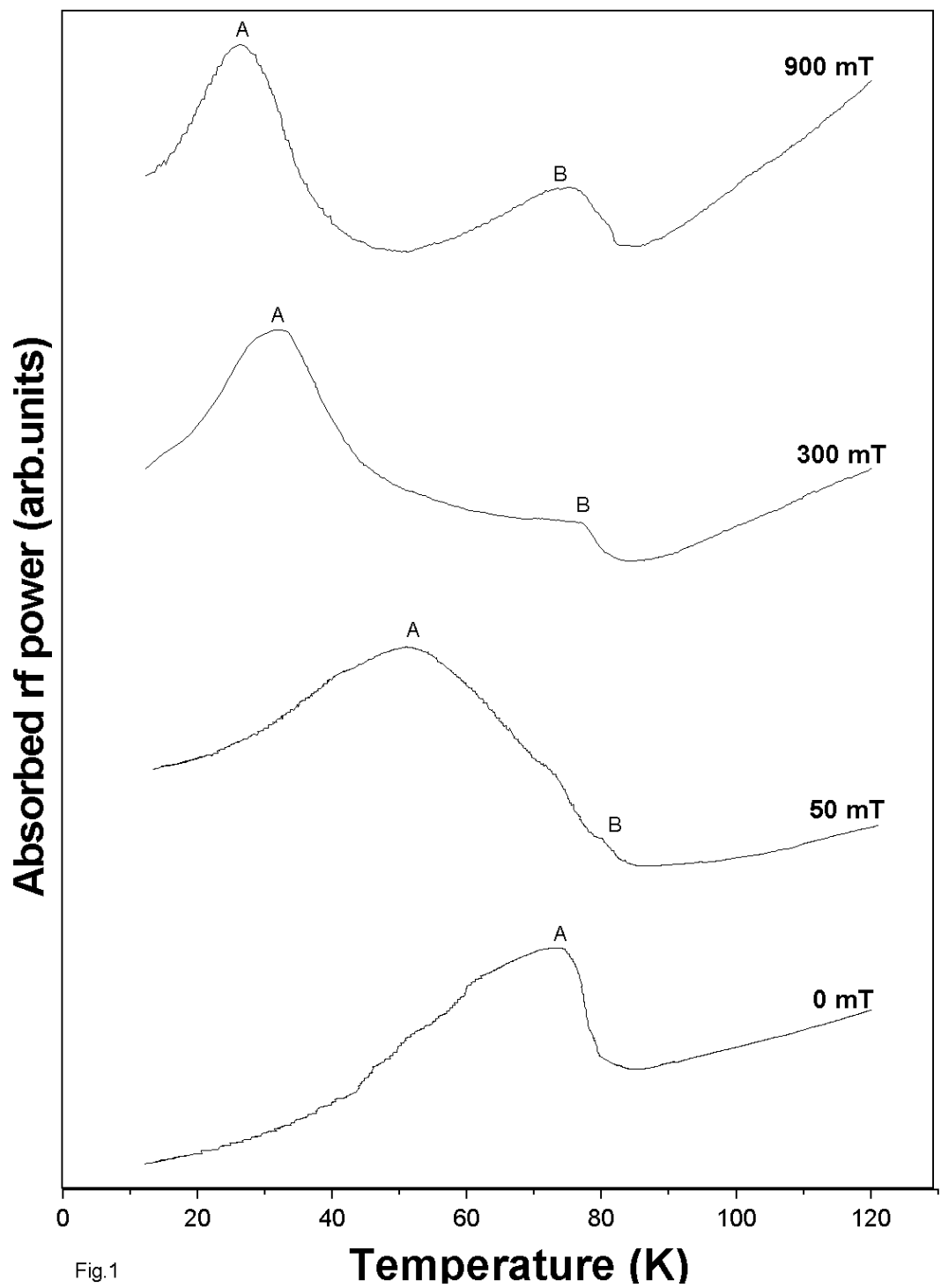

Fig.1

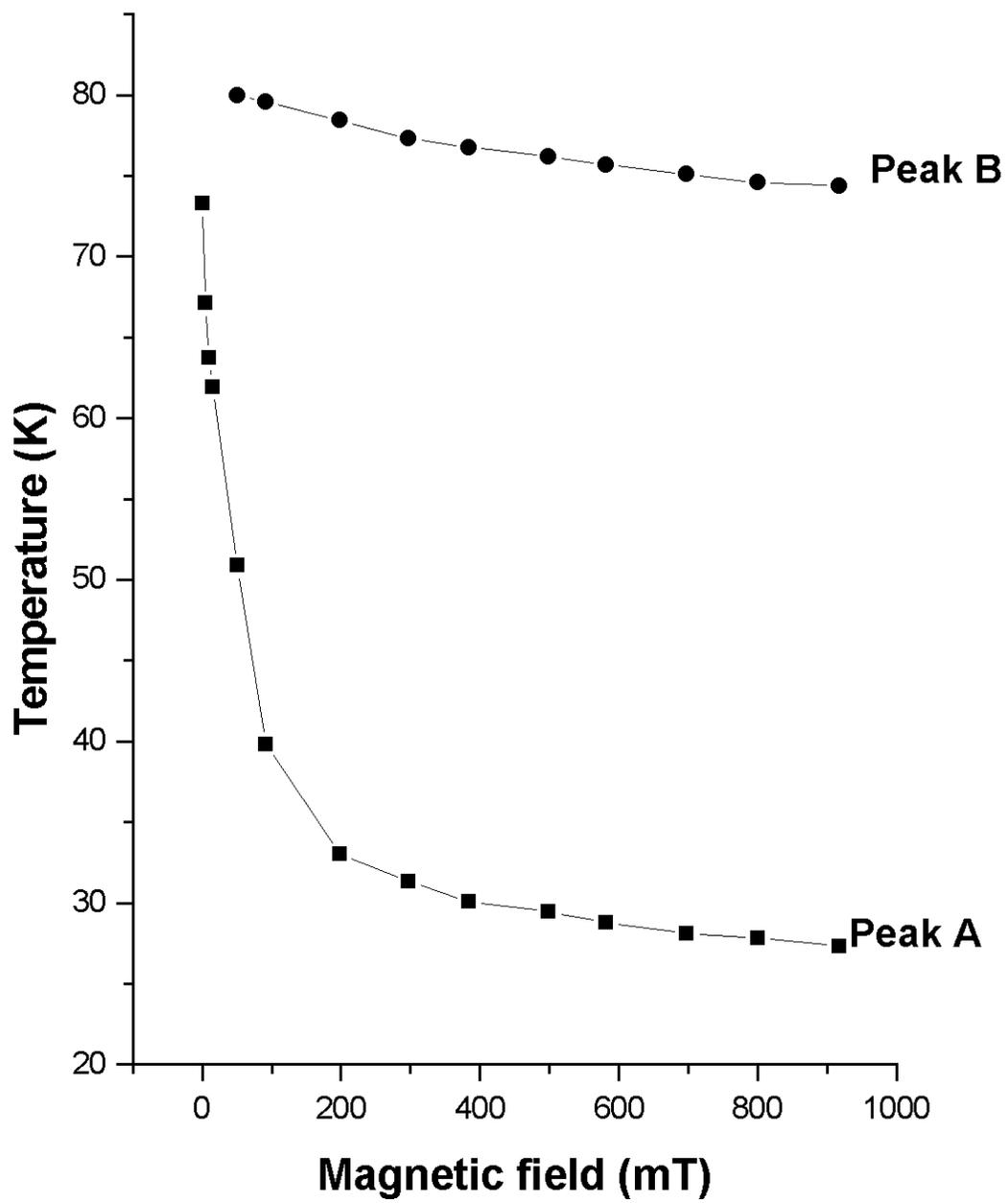

Fig.2

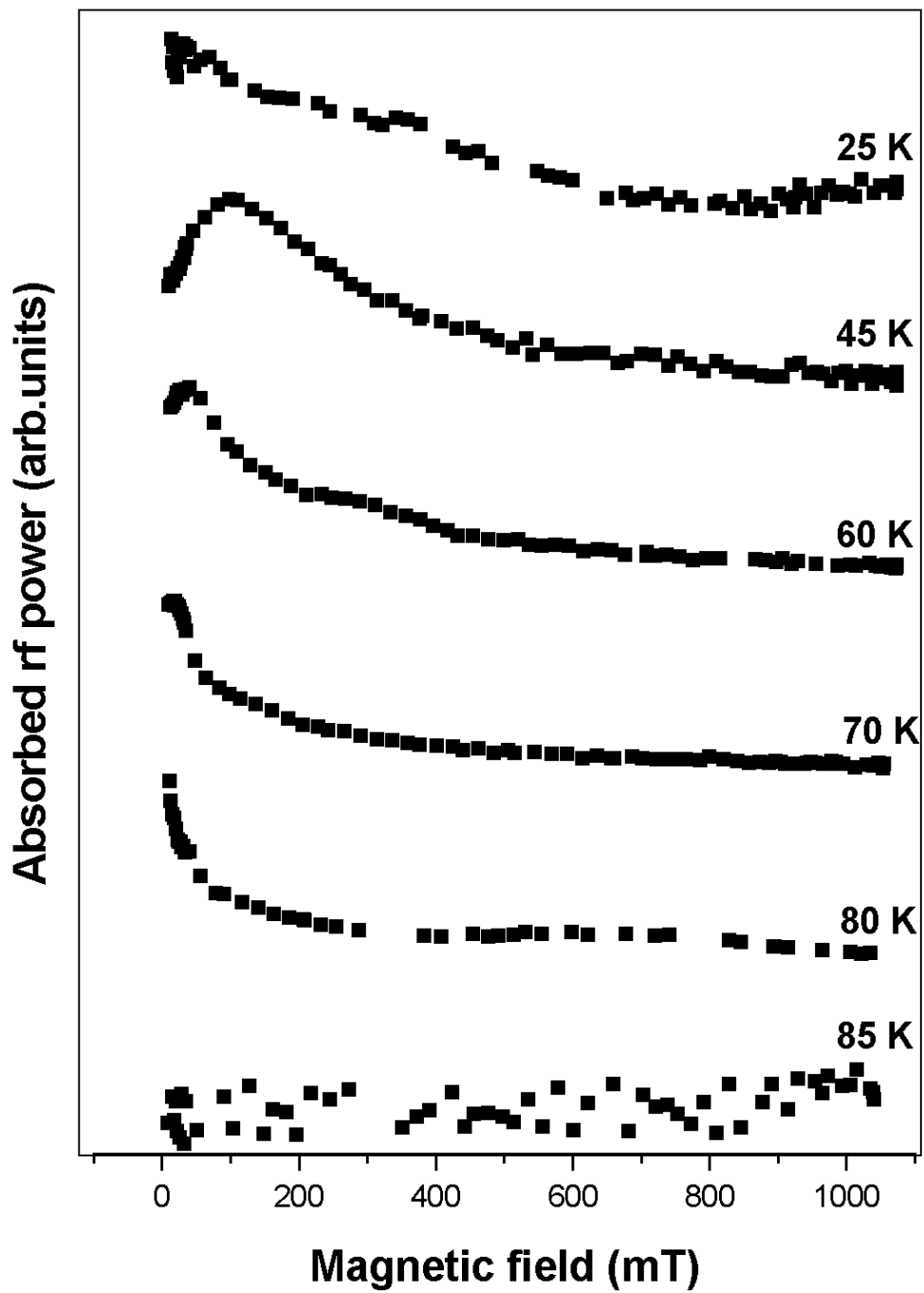

Fig.3

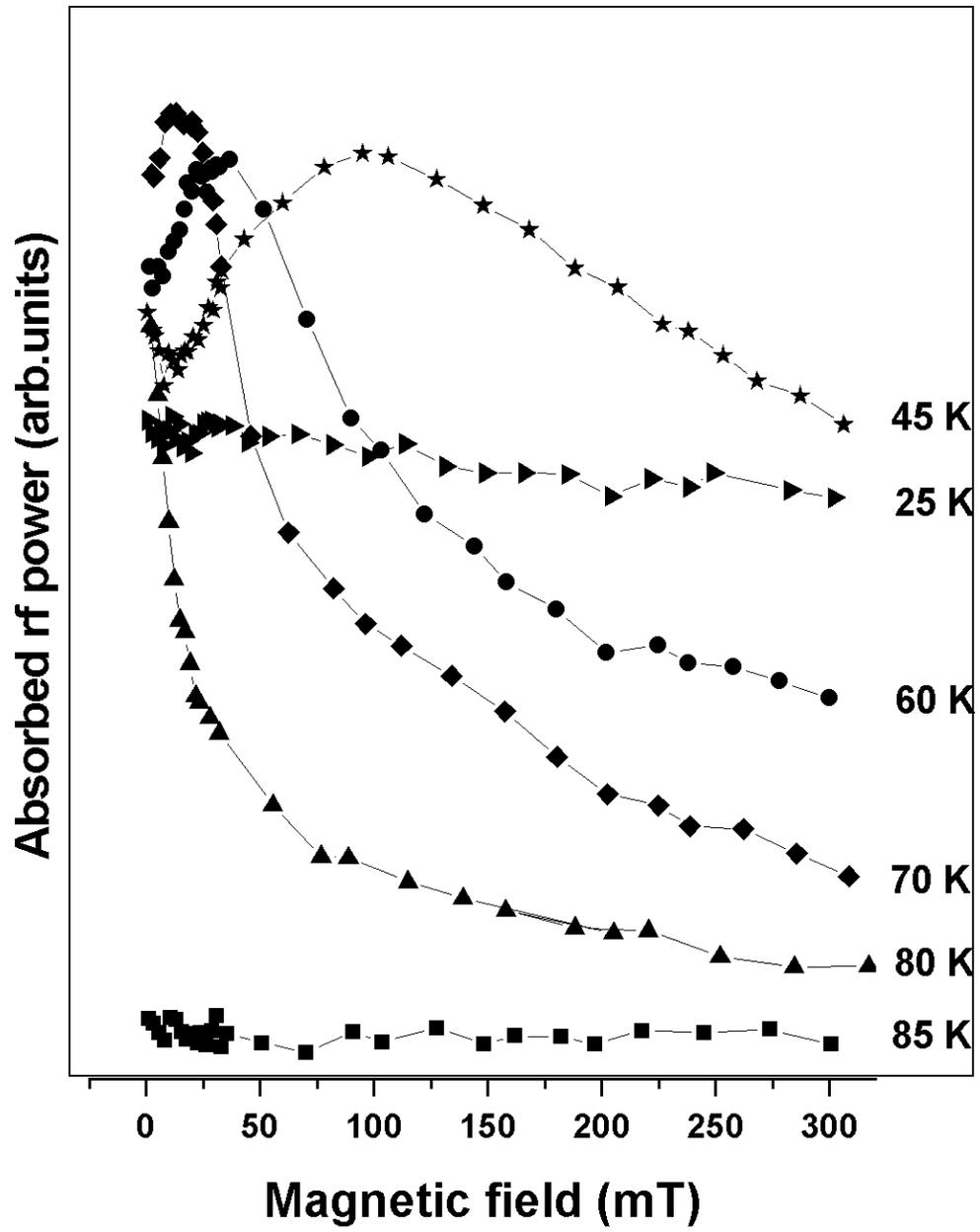

Fig.4